\documentclass[a4paper]{article}
\usepackage[ansinew]{inputenc}
\usepackage{amssymb}
\usepackage{graphicx}
\usepackage{amsmath}
\def\eq#1{$$#1$$}

\begin{document}
\title{Quantum key distribution over 30\,km of standard fiber using energy-time entangled photon pairs: a comparison of two chromatic dispersion reduction methods}
\author{Sylvain Fasel, Nicolas Gisin, Grégoire Ribordy, Hugo Zbinden\\
Group of Applied Physics, University of Geneva\\
20, rue de l'École-de-Médecine\\
CH-1211 Geneva 4, Switzerland
}
\maketitle
\begin{abstract}
We present a full implementation of a quantum key distribution system using energy-time entangled photon pairs and functioning with a 30\,km standard telecom fiber quantum channel. Two bases of two orthogonal states are implemented and the setup is quite robust to environmental constraints such as temperature variation. Two different ways to manage chromatic dispersion in the quantum channel are discussed.
\end{abstract}
Since the birth of quantum key distribution (QKD) \cite{bb84,ekert,ekert2}, a lot of research and discoveries have been made \cite{townsend,merolla,jmodopt,sargpra} leading today to demonstrated long distance QKD \cite{hughes,stucki,tomita,shields} and even to commercially available quantum cryptography devices. However, these systems rely on faint laser pulses containing about 0.1 photon per pulse to guarantee absolutely secure keys. Therefore, only a fraction of the pulses will lead to effective bit transmission. Another consequence is that more than one photon are present per pulse with non-zero probability. True single photon sources \cite{singlephoton,singlephoton2} seem to be a promising solution, but existing devices are not yet usable for out-of-the-lab systems. The most serious alternatives are systems based on entangled photon pairs \cite{zeilinger,kwiat,tittel,ribordy}, but no real-world long distance QKD using this approach has been reported yet, mainly for two reasons. First, the setup itself is relatively complex. Second, the spectral characteristics of the photon pairs sources imply that chromatic and polarization dispersion effects are not negligible and increase the error rate. Consequently, one has to find solutions if the aim is to deploy systems over fibre telecom networks.\\
The goal of the present article is to demonstrate solutions to the chromatic dispersion issue for an energy-time entanglement based QKD system using a standard fiber quantum channel. We improved a setup previously presented by our group \cite{ribordy} by using dispersion compensation or spectral filtering and extended the transmission range.

\begin{figure}[htbp]
\includegraphics[width=\columnwidth]{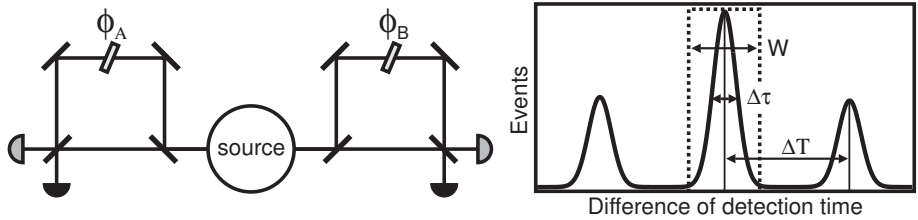}
\caption{Schematic diagram of a Franson-type interferometers-source arrangement and the corresponding event frequency plotted as a function of the difference of detection time at Alice and at Bob. The resulting peaks can be approximated by Gaussian with FWHM $\Delta\tau$ and are separated by a time $\Delta T$ corresponding to the path length difference. A window discriminator of width $W$ is placed around the central peak. The two phases $\phi_A$ and $\phi_B$ modulate the  correlations.}
\label{fransonsmall}
\end{figure}

Our setup is based on a Franson arrangement \cite{franson} of interferometers. A parametric down conversion photons pair source is located between Alice and Bob. They both have an unbalanced Mach-Zender interferometer with photon-counting detectors connected at all outputs. When considering a given photon pair, four different events can be detected by both Alice and Bob. First, the photons can both propagate through the short arms of the interferometers. Then, one can take the long arm at Alice while the other takes the short one at Bob. The opposite is also possible. Finally, both photon can propagate through the long arms. When the path differences of the interferometers are matched within a fraction of the coherence length of the down-converted photons, the short-short and the long-long processes are indistinguishable and thus yield two-photons interferences, provided that the coherence length of the pump photons is longer than the path difference. If one records events as a function of time difference between detections at Alice and at Bob, 3 peaks appear (figure \ref{fransonsmall}). The central one is constituted by the interfering short-short and long-long events. It can be distinguished from the others by placing a time window discriminator, and is used to isolate nonlocal quantum correlation between Alice's and Bob's detections.
On figure \ref{fransonsmall} we see that to allow the window discriminator to take the maximum of the central peak into account (to have an optimal detection rate) but at the same time avoid the side peaks (non-correlated side peaks detections introduce errors), the separation between the peaks $\Delta T$ must be significantly bigger than $\Delta\tau$. $\Delta\tau$ is the RMS of all the temporal spreading contributions: photon's coherence time, electronic and detection jitter and, for dispersive medium between the source and the interferometers, the spreading of the wave packet due to chromatic dispersion. This condition reads:
 \eq{
 \Delta\tau<\Delta T<t_c^{pump}
 }
where $t_c^{pump}$ is the coherence time of the pump laser.\\
Dispersive media introduce chromatic dispersion of $D=\frac{\delta\tau}{\delta\lambda}$ per unit length. For standard telecom fibers and light around 1550\,nm this value is $D\cong17\,\textrm{ps}\,\textrm{nm}^{-1}\textrm{km}^{-1}$. The gaussian uncertainty in time due to chromatic dispersion only is then given by $\Delta\tau_{\textrm{disp}}=D\,\Delta\lambda\,L$ (where $\Delta\lambda$ is the photon's spectral width and $L$ the length of the fiber). Consequently, the inequality
\eq{
D\,\Delta\lambda\, L<\Delta T
}
has to be fulfilled to perform QKD with negligible errors due to chromatic dispersion. To match this condition for a given length of fiber, one can use large $\Delta T$, reduce the dispersion $D$, or use photons with fine spectral width. We do not have to deal with similar problems caused by polarization mode dispersion, as information is not encoded in polarization.

Theoretically, $\Delta T$ could be as large as needed by increasing the path length difference of the interferometers. For example, a  difference larger than 3\,m is needed for 30\,km of standard fiber with our source. However, interferometers of this size are hard to stabilize. Moreover fiber interferometers suffer from a chromatic dispersion difference between the 2 arms that reduces the visibility. Another point is that even if there is no overlap between the peaks, dispersion broaden them. Consequently, wider coincidence windows are required, increasing the error contribution of the detectors' noise. For theses reasons we did not implement this solution.\\
In order to reduce $D$ it is possible to use dispersion shifted fiber for the quantum channel as in \cite{ribordy}. This solution is not relevant because the resulting setup would not be deployable in a real-world telecom network, as there are only few installed lines using this kind of fibers.\\
Another possibility is to use dispersion compensation \cite{compens}, with the only drawback of the added loss on the quantum channel (which lowers the signal over noise ratio). This solution is the first that we investigate in this article.\\
Alternatively, we reduce the $\Delta\lambda$ of the 1550\,nm photons (for 30\,km the above necessary condition implies $\Delta\lambda<5.5$\,nm) by inserting a bandpass filter on Alice's side. As the central wavelength and the spectral width of the 2 down converted photons are related by energy conservation, filtering on one side reduces the spectral width of the twin photons detected on the other side. Bob's detectors are gated upon Alice's detections thus, even if the key rate is reduced, the signal over noise ratio remains constant.

To implement the full BB84 protocol, two different measurement bases are needed. This can be done by using 2 interferometers or a fast switch inside a single interferometer. Here we use a birefringent interferometer where the phase applied on the photons depends on the polarization. Thanks to this, we implement the full energy-time entangled BB84 protocol \cite{ekert,ekert2} with only 2 interferometers instead of 4. More details can be found in \cite{ribordy}.
\begin{figure}[htbp]
\includegraphics[width=\columnwidth]{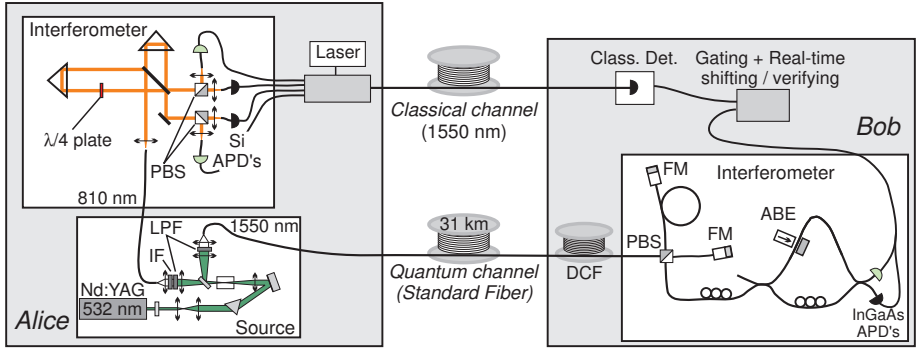}
\caption{Schematic diagram of the experimental setup (PBS: polarizing beam splitter; LPF: low-pass pump filters; IF: optional 2\,nm bandpass filter; DCF: optional dispersion compensating fiber; FM: Faraday mirrors; ABE: adjustable birefringent element used as fibre $\lambda/4$ plate)}
\label{schema}
\end{figure}
Pairs of 810\,nm--1550\,nm photons are produced from a type I configuration of $\textrm{KNbO}_3$ crystal pumped by a 532\,nm continuous laser. The 810\,nm photons are collected in a single mode fiber and sent to the Alice's interferometer which is made of bulk optical components. It has 4 outputs, each of them consisting of a single mode fiber coupling system followed by a passively quenched silicium photon counter (EG\&G). The 1550\,nm photons are collected in a single mode fiber and launched into the quantum channel which is a spool of several kilometers of standard fiber. It can be optionally followed by our dispersion compensation device, i.e. a spool of negative dispersion fiber. The Bob's interferometers is connected at the other side of the quantum channel, together with a polarization beam-splitter and Faraday mirrors system that are part of the BB84 implementation. Two gated-mode InGaAs detectors (idQuantique) are connected at the interferometer's outputs. The path difference of both interferometers correspond to  1\,m optical length, and consequently $\Delta T\cong$3.3\,ns. \\
An electronic system is used to trig Bob's detectors whenever a photon is detected by any of Alice's detectors. This electronics can also be used to characterize the system in real-time: the information about which of the Alice's detectors registered a count can optionally be coded and sent to Bob's through a synchronized classical channel (consisting of another spool of standard fiber) using a 1550\,nm laser. Upon detection at Bob's side, this information can be used for sifting and to verify the bits. This electronics thus enables to have an immediate information about the sifted raw key creation rate and the error rate.\\ 
The most relevant experimental parameters are the following: spectral FWHM of the down converted photons: 6.9\,nm at 1550\,nm, 2\,nm at 810\,nm; probability of having a photon coupled into the quantum channel whenever the 810\,nm silicium detector fires: 0.5; losses of Bob's apparatus: 5.4\,dB; quantum efficiency of Bob's detectors: 10\%; detection gate width $W=1.1$\,ns; false counts per gate due to the average noise of both detectors at Bob: $\sim1.13\times10^{-5}$; false counts per gate due to noise on Alice's side: negligible;\\
The whole setup requires  very careful tuning. The most difficult part is the alignement of Alice's bulk interferometer. In particular, we have to pay attention that the probabilities for the short-short and long-long events are equals for all combination of detections at Alice and at Bob \cite{entangconcentration}. Bob's apparatus must also be aligned to ensure that the polarization transformations in both arms of the interferometer are identical.

We used a quantum channel consisting of a spool of 31\,km of standard fiber, inducing losses of 8.3\,dB. Two different configurations were implemented to limit the effect of chromatic dispersion:\\
In the first one, the dispersion compensating spool (OFS) was connected in between the quantum channel and Bob's apparatus. This device compensate a dispersion of $D_{comp}\cong506\,\textrm{ps}\,\textrm{nm}^{-1}$, which corresponds to about 30\,km of standard fiber. It induces losses of $2.9$\,dB, and a delay corresponding to 4\,km of standard fiber. A 31+4\,km spool of standard fiber was consequently used for the classical channel. In this case we implemented the full two bases BB84 protocol using 4 detectors on Alice's side. The total count rate at Alice was about 79\,kHz. The visibility of the interferences was about 89\%.\\
In the second configuration, an interferometric bandpass filter was placed in the source apparatus just before the collection lenses of the 810\,nm output port (see IF on fig. \ref{schema}). We used a filter of 2\,nm FWHM centered at 814\,nm (as this has been determined to be optimal for production/collection) and consequently the central wavelength of the co-detected photons was 1536\,nm. As the Gaussian 2\,nm filter acts on an approximately 2\,nm wide Gaussian spectrum, we obtained a width of about 1.45\,nm FWHM at 814\,nm, and 5.2\,nm at 1536\,nm. The number of 810\,nm photons detected was reduced by a factor of about 3. To facilitate the alignements and the measurements in this second configuration, we increased the coincidences count rate by using only one of the two BB84 bases. The total count rate at Alice was about 36\,kHz and the visibility of the interference was about 92\% in this case.

\begin{figure}[htbp]
\includegraphics[width=0.5\columnwidth]{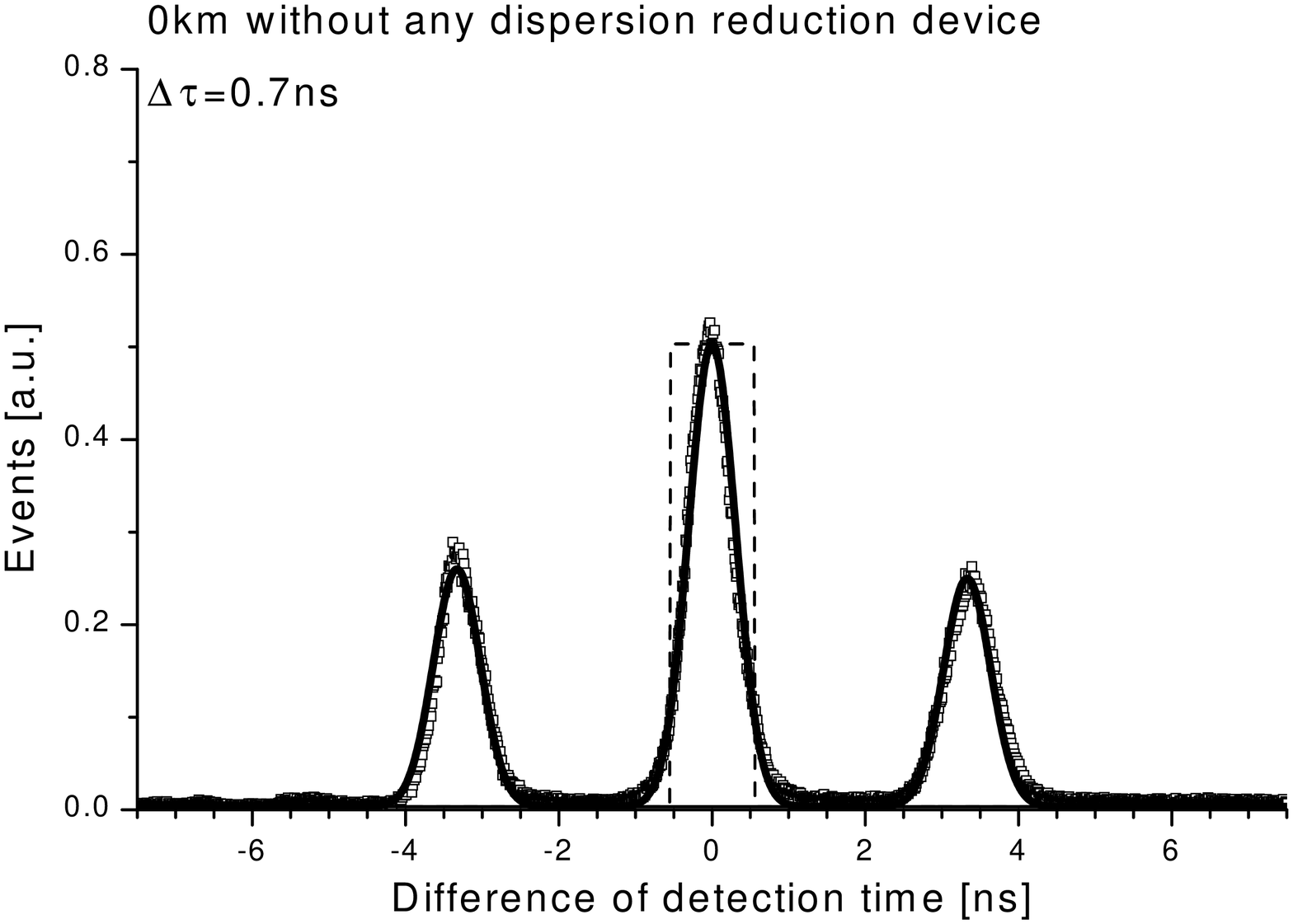}
\includegraphics[width=0.5\columnwidth]{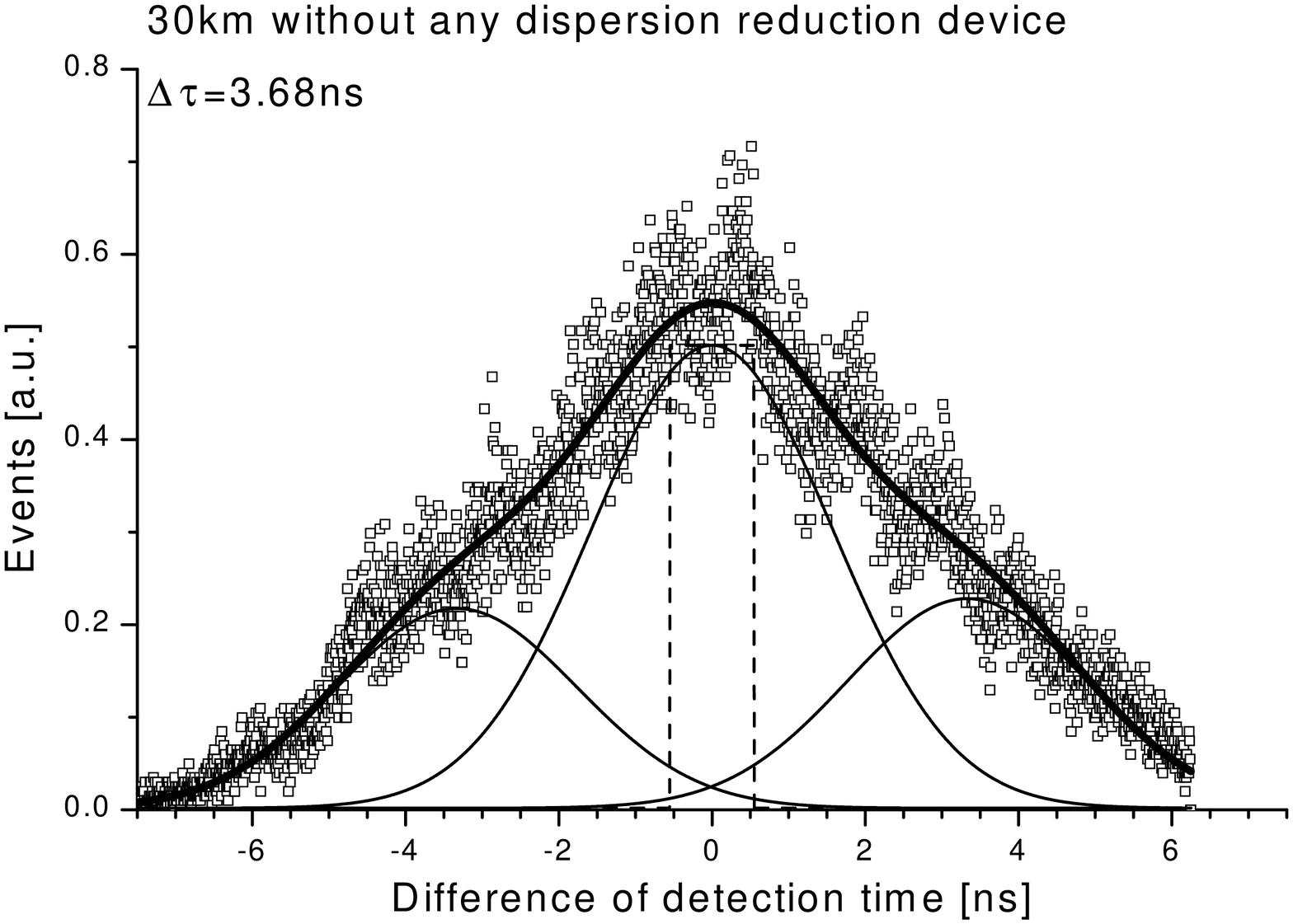}
\includegraphics[width=0.5\columnwidth]{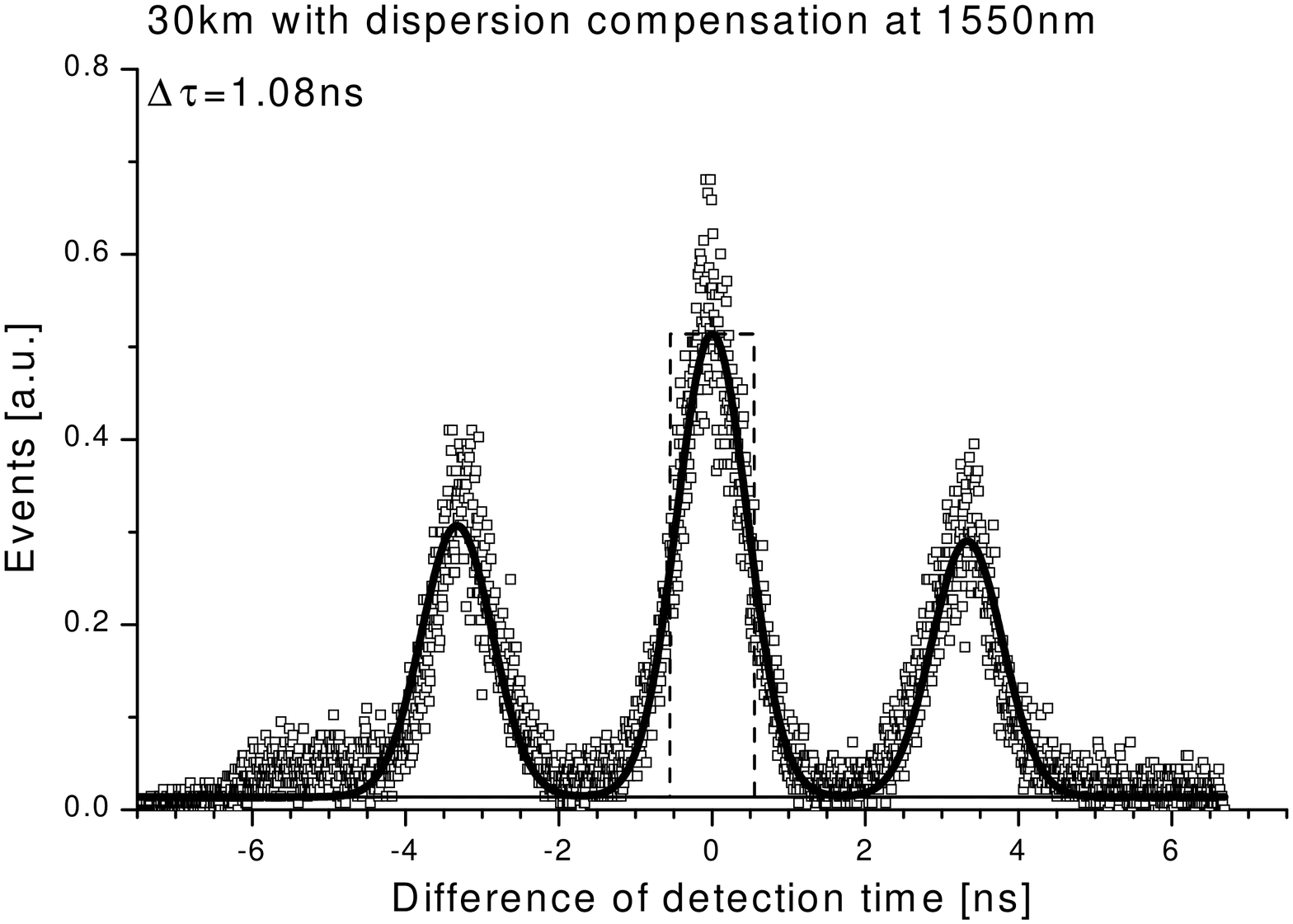}
\includegraphics[width=0.5\columnwidth]{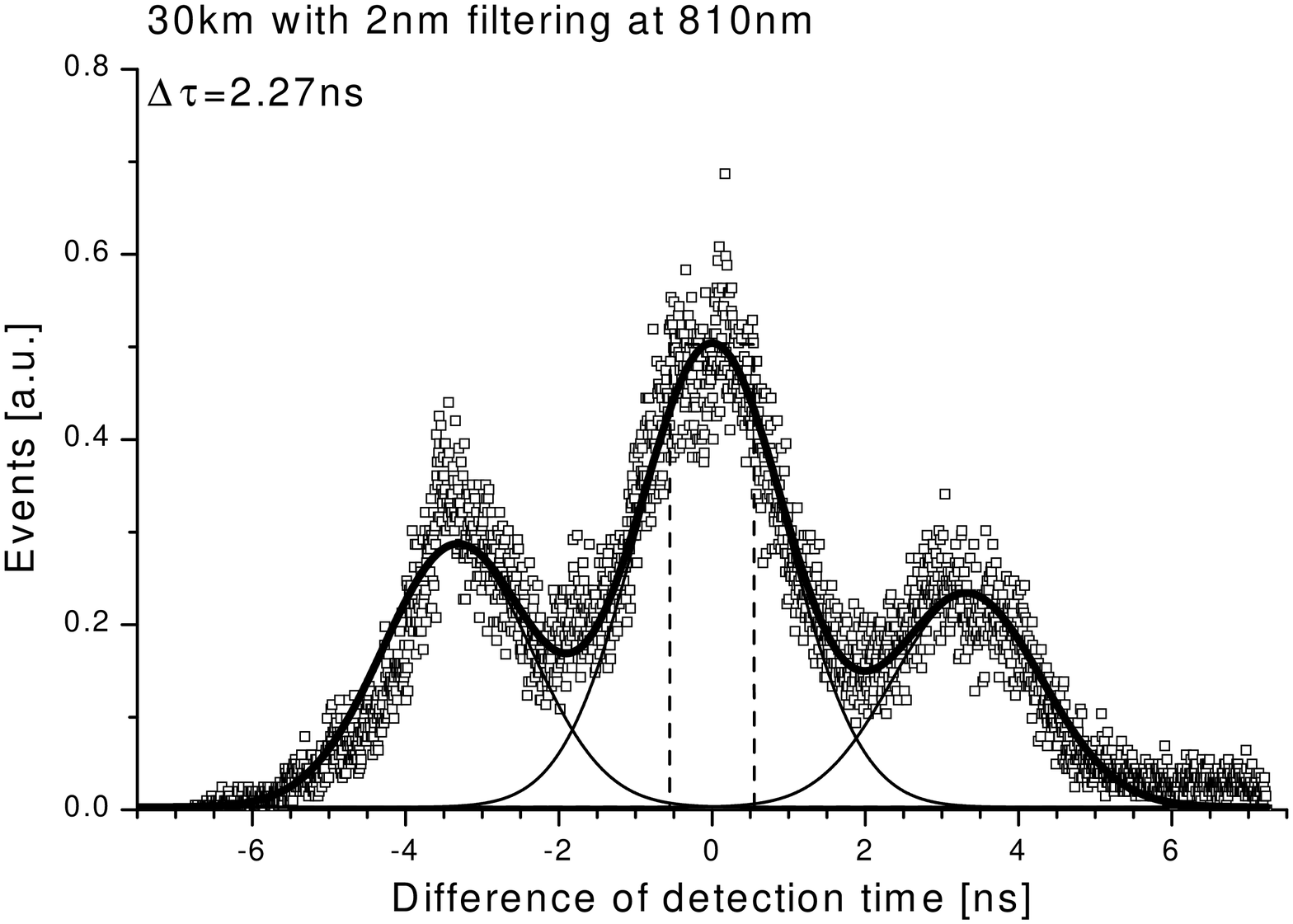}
\caption{Distribution of coincidence detection times for different setup. The squares show the experimental data. The plain curves show the individual gaussian peak; the heavy curve shows the sum of the 3 peak's contributions which is fitted to the experimental data to extract the FWHM $\Delta\tau$ of the peaks; the dashed lines show the position and width of the detection gate.\label{3peaks}}
\end{figure}

Figure \ref{3peaks} shows the resulting temporal distributions of the events obtained for 4 setups. Using the first plot, we obtained the total electronic/detection jitter of value 0.7\,ns, because it is directly equal to the FWHM of the (central) peak; we also use this plot to verify that $\Delta T=3.3\,$ns. The second plot clearly shows that a dispersion reduction method is necessary. Indeed, in the absence of any dispersion reduction, the two side peaks overlap to a large extend the detection window. We determined that in this case, the error rate due to the only contribution of uncorrelated events counted inside the detection windows is already about 10\% of the total count rate. Moreover, the peaks are more than 3 time larger than the detection windows, lowering the detection rate. The effect of the dispersion reduction is clearly visible on the third and fourth plots, corresponding to the two configurations described above. Note that the available filter of 2\,nm does not completely remove the effect of chromatic dispersion. However, a narrower filter would decrease even more the total count rate.

We achieved key distribution for our solutions. Because of phase instabilities in the interferometers the duration of a key exchange was limited to about 40-50 minutes, but this issue could be addressed by using actively stabilized interferometers \cite{stabili}. During this period the quantum bit error rate (QBER), the ratio of the error rate over the total rate after sifting, was about 10\% on average. Table \ref{qbers} summarizes the performance obtained in terms of the different sources of errors.
\begin{table}[htbp]
\begin{center}
\caption{QKD performances for both implementations}
\label{qbers}
\begin{tabular}{lcccccc}
\hline
\hline
Configuration&Sifted&Opt.&Acc.&Detect.&Disp.&Tot.\\
             &key rate&QBER&QBER&QBER&QBER&QBER\\
\hline
compensation&23\,Hz           &5.5\%& 1\%&   4\%& $\cong$0\% & 10.5\%\\
filtering   &12\,Hz&          4\%&  1\% &   1.7\%& 0.5\% &       7.2\%\\
\hline\hline
\end{tabular}
\end{center}
\end{table}
The total QBER  is the sum of several contributions: i) The optical error rate due to the imperfect contrast of the interferences. ii) The accidental coincidences due to the non-zero probability of creating two pairs during the detection gate time-width: the photons from two different pairs are not entangled and thus have a 50\% probability of producing incorrect bits. iii) The detectors' noise, which is independent of distance. As the detection count decrease with the losses in the line this detector QBER contribution increases with the distance and is generally the main contribution for large distances. iv) The dispersion which makes some non-correlated events to be registered inside the detection gate, as explained above. To estimate this QBER contribution we used the measurement presented in figure \ref{3peaks}. Using the fitted $\Delta\tau$ of the peaks, it is possible to numerically integrate the 3 gaussian curves inside the discriminator window to obtain the part of detections which comes from the side peaks. This is the error rate arising from the chromatic dispersion effect. Note that, for given $\Delta T$ and $W$, this integral value grows rapidly with $\Delta\tau$ when this variable reaches about one third of $\Delta T$. The chromatic dispersion QBER is thus very sensitive to spectral width and fiber length.\\
The detector QBER is more important in the compensation configuration mainly because the compensating device add losses on the quantum channel, reducing the signal over noise ratio by a factor 2. However, one of our Bob's detectors was about 5 times less noisy than the other. With two such detectors, the detection QBER for the compensating solution would drop below 1\%.\\
The numerical integration of the events peaks was also used to calculate the fraction of the incoming photons which are accounted for bits of key. If all the photons that are part of the short-short/long-long events were inside the detection gate this number would be 0.5. In our case it is 0.38 for the compensation solution and 0.22 for the filtering solution as the events are more widely distributed in the second case (see figure \ref{3peaks}). These values are taken into account as a small added loss for the filtering solution with respect to the compensating one. 
\begin{table}[htbp]
\begin{center}
\caption{Factors leading to sifted key rates}
\label{losstable}
\begin{tabular}{lccccccccc}
\hline
\hline
Configuration&Singles&   $\mu$&$T_L$&$T_B$&$T_C$&$\eta_d$&$\eta_g$&$q_s$&Sifted\\
             &rate&    & dB    & dB    &  dB   &        &        &     &key rate\\
\hline
compensation&79\,kHz           &0.5&8.3&5.4&2.9&0.1&0.38&0.7&23\,Hz\\
filtering   &36\,kHz          &0.5&8.3&5.4&0&0.1&0.22&0.7&12\,Hz\\
\hline\hline
\end{tabular}
\end{center}
\end{table}
Table \ref{losstable} summarizes the different factors leading to the registered sifted key rates, starting from the singles rate at Alice. These factors are: $\mu$: probability of having a photon coupled into the quantum channel whenever the 810\,nm silicium detector fires; $T_L$: loss of the quantum channel; $T_B$: loss of Bob's apparatus; $T_C$: loss of the dispersion compensating fiber spool; $\eta_d$: quantum efficiency of Bob's detectors; $\eta_g$: fraction of the incoming photons which are accounted for bits; $q_s$: proportion of the bits that remain after sifting. This number should be 0.5 for a perfectly balanced two bases system. In the compensation case the value 0.7 is explained by two reasons. First, the bases choice at Alice is biased by the differences between the four bulk-to-fiber coupling and detectors efficiencies.  Secondly, the passive choice of bases at Bob depends on the polarization of the incoming photons. These photons are only partially depolarized by 30\,km of fiber, and the remaining polarization fluctuate inside the quantum channel during the key exchange. In the filtering case, as we use only one base at Alice, we could achieve a value of 1 by tuning the passive choice at Bob if the incoming photons where perfectly polarized, but this is not the case. Moreover, as in the compensation case, the polarization fluctuates inside the quantum channel. Note that a $q_s\neq0.5$ does not impair on the security of the scheme \cite{asymetricbases}.

From theses results, we see that the choice of the best suitable dispersion reduction method is a matter of tradeoff between QBER and key rate values, and is different for each particular setup. In our case, numerical estimations show that an optimized compensation solution is better in term of key rate beyond 15\,km, for a given QBER. However, the amount of negative dispersion introduced must be calculated specifically for a given length of the quantum channel fiber. The main practical advantage of the filtering solution is thus that the system can be uniquely designed to be usable over a wide range of distances.  The resulting key rate decrease can be compensated by pumping the source with a more powerful laser, or by using a more efficient photon pair source such as periodically poled lithium niobate waveguide \cite{sebppln}. When this is possible, filtering is more useful for real application. The only problem resulting from a configuration using a heavily pumped source filtered at Alice is the increase of accidental uncorrelated coincidence counts. This issue can be solved by filtering the photons at Bob with a corresponding filter. As both wavelengths are correlated, the only drawback is the non unity peak transmission of the filter. 

\begin{figure}[h]
\begin{center}
\vspace{0.5cm}
\includegraphics[width=0.75\columnwidth]{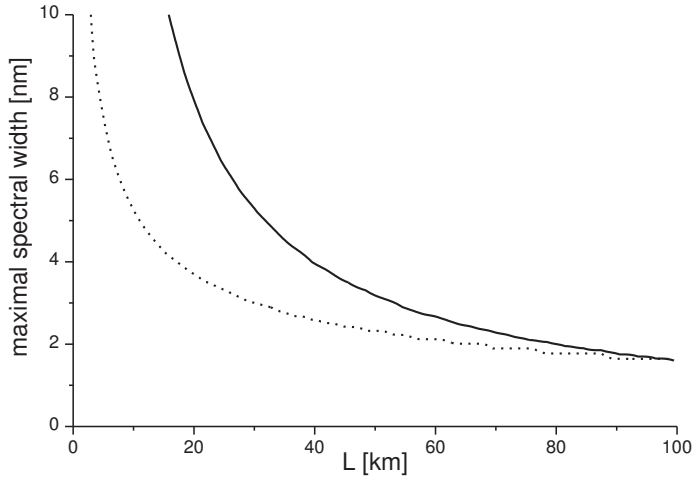}
\caption{Maximal spectral width that keeps chromatic and polarization dispersion induced QBER below 1\% as function of the quantum channel length. Plain curve: our energy-time entanglement implementation, with $\Delta T=3.3$\,ns; dotted curve: possible optimal polarization entanglement implementation with PMD values of 0.1\,ps\,km$^{-\frac12}$}
\label{PMDvsET}
\end{center}
\end{figure}
The spectral width of the photons is also a limiting factor for polarization entanglement based QKD. QBER increasing in the case of polarization QKD is due to polarization mode dispersion (PMD) that depolarizes low coherent photons. Figure \ref{PMDvsET} shows the calculated maximal spectral width of 1550\,nm photons that keeps the chromatic/polarization dispersion induced QBER below 1\%. The curve for energy-time entangled photons is calculated using our particular experimental parameters, while the one for polarization is calculated using $QBER=0.5\times(1-\overline{DOP})$ where $\overline{DOP}$ is the average degree of polarization computed numerically from formulas developed in \cite{PMDgisin}, using the standard PMD value of 0.1\,ps\,km$^{-\frac12}$. We see that for distances up to about 100\,km, the energy-time solution is more robust to large spectral width. For longer ranges both solutions become similar with a slight advantage for the one using polarization, because of the square root dependance of the PMD with distance. However, these curves strictly apply to dispersion QBER and do not take into account technical difficulties related to the necessary active polarization state control.

In this article, we presented two practical means of dealing with chromatic dispersion induced problems in long-range QKD using entangled photon-pairs: dispersion compensation and reduction of the photons' spectral width at Alice. Optimal parameters where investigated and these solution were demonstrated by implementing a partial and complete BB84-like protocol with a setup featuring characteristic that could lead to a real-world telecom applications. A secret key was distributed over 30\,km of standard fiber at a sifted bit rate of more than 20\,Hz and with an average QBER below the 11\% limit for absolute security as stated in \cite{preskill_shor}. Higher key rate is possible using more efficient sources. However, for a practical implementation, actively stabilized interferometers are needed.

The authors would like to thank Claudio Barreiro and Jean-Daniel
Gautier for technical support. Financial support by the Swiss NCCR
Quantum Photonics is acknowledged.

\end{document}